\documentclass[11pt]{article}
\usepackage{graphicx}
\usepackage{dcolumn}     
\usepackage{bm}

\newcommand{\BABARPubYear}    {04}
\newcommand{\BABARConfNumber} {21}
\newcommand{\SLACPubNumber} {10618}

\input babarsym

\long\def\inst#1{\par\nobreak\kern 4pt\nobreak
    {\it #1}\par\vskip 10pt plus 3pt minus 3pt}

\RequirePackage{xspace}

\usepackage{relsize}
\def\babar{\mbox{\slshape B\kern-0.1em{\smaller A}\kern-0.1em
    B\kern-0.1em{\smaller A\kern-0.2em R}}}

\def\en         {\ensuremath{e^-}\xspace}   
\def\ep         {\ensuremath{e^+}\xspace}
 
\def\epem       {\ensuremath{e^+e^-}\xspace}

\def\mup        {\ensuremath{\mu^+}\xspace}
\def\mun        {\ensuremath{\mu^-}\xspace} 
\def\mumu       {\ensuremath{\mu^+\mu^-}\xspace}

\def\taup       {\ensuremath{\tau^+}\xspace}
\def\taum       {\ensuremath{\tau^-}\xspace}

\def\ellell     {\ensuremath{\ell^+ \ell^-}\xspace}

\def\nub        {\ensuremath{\overline{\nu}}\xspace}
\def\nunub      {\ensuremath{\nu{\overline{\nu}}}\xspace}
\def\nub        {\ensuremath{\overline{\nu}}\xspace}
\def\nunub      {\ensuremath{\nu{\overline{\nu}}}\xspace}
\def\nue        {\ensuremath{\nu_e}\xspace}
\def\nueb       {\ensuremath{\nub_e}\xspace}

\def\num        {\ensuremath{\nu_\mu}\xspace}
\def\numb       {\ensuremath{\nub_\mu}\xspace}

\def\nut        {\ensuremath{\nu_\tau}\xspace}
\def\nutb       {\ensuremath{\nub_\tau}\xspace}

\def\nul        {\ensuremath{\nu_\ell}\xspace}
\def\nulb       {\ensuremath{\nub_\ell}\xspace}

\def\g     {\ensuremath{\gamma}\xspace}
\def\gaga  {\ensuremath{\gamma\gamma}\xspace}  
\def\ggstar{\ensuremath{\gamma\gamma^*}\xspace}

\def\qqbar {\ensuremath{q\overline q}\xspace}

\def\piz   {\ensuremath{\pi^0}\xspace}

\def\ppz   {\ensuremath{\pi^0\pi^0}\xspace}
\def\pip   {\ensuremath{\pi^+}\xspace}
\def\pim   {\ensuremath{\pi^-}\xspace}
\def\pipi  {\ensuremath{\pi^+\pi^-}\xspace}

\def\pimp  {\ensuremath{\pi^\mp}\xspace}

\def\kaon  {\ensuremath{K}\xspace}
\def\Kbar  {\kern 0.2em\overline{\kern -0.2em K}{}\xspace}

\def\Kz    {\ensuremath{K^0}\xspace}
\def\Kzb   {\ensuremath{\Kbar^0}\xspace}
\def\KzKzb {\ensuremath{\Kz \kern -0.16em \Kzb}\xspace}
\def\Kp    {\ensuremath{K^+}\xspace}
\def\Km    {\ensuremath{K^-}\xspace}
\def\Kpm   {\ensuremath{K^\pm}\xspace}

\def\KpKm  {\ensuremath{\Kp \kern -0.16em \Km}\xspace}
\def\KS    {\ensuremath{K^0_{\scriptscriptstyle S}}\xspace} 
\def\KL    {\ensuremath{K^0_{\scriptscriptstyle L}}\xspace} 
\def\Kstarz  {\ensuremath{K^{*0}}\xspace}

\def\Kstar   {\ensuremath{K^*}\xspace}

\def\Kstarp  {\ensuremath{K^{*+}}\xspace}

\def\Dbar    {\kern 0.2em\overline{\kern -0.2em D}{}\xspace}

\def\Dz      {\ensuremath{D^0}\xspace}
\def\Dzb     {\ensuremath{\Dbar^0}\xspace}
\def\DzDzb   {\ensuremath{\Dz {\kern -0.16em \Dzb}}\xspace}
\def\Dp      {\ensuremath{D^+}\xspace}
\def\Dm      {\ensuremath{D^-}\xspace}

\def\DpDm    {\ensuremath{\Dp {\kern -0.16em \Dm}}\xspace}
\def\Dstar   {\ensuremath{D^*}\xspace}

\def\Dstarz  {\ensuremath{D^{*0}}\xspace}

\def\Dstarp  {\ensuremath{D^{*+}}\xspace}
\def\Dstarm  {\ensuremath{D^{*-}}\xspace}

\def\B       {\ensuremath{B}\xspace}
\def\Bbar    {\kern 0.18em\overline{\kern -0.18em B}{}\xspace}
\def\Bb      {\ensuremath{\Bbar}\xspace}
\def\BB      {\ensuremath{\B {\kern -0.16em \Bb}}\xspace}
\def\Bz      {\ensuremath{B^0}\xspace}
\def\Bzb     {\ensuremath{\Bbar^0}\xspace}
\def\BzBzb   {\ensuremath{\Bz {\kern -0.16em \Bzb}}\xspace}
\def\NBB     {\ensuremath{N_{\BB}}}

\def\Bu      {\ensuremath{B^+}\xspace}
\def\Bub     {\ensuremath{B^-}\xspace}

\def\Bpm     {\ensuremath{B^\pm}\xspace}

\def\BpBm    {\ensuremath{\Bu {\kern -0.16em \Bub}}\xspace}

\def\BorBbar    {\kern 0.18em\optbar{\kern -0.18em B}{}\xspace}
\def\DorDbar    {\kern 0.18em\optbar{\kern -0.18em D}{}\xspace}
\def\KorKbar    {\kern 0.18em\optbar{\kern -0.18em K}{}\xspace}

\def\jpsi     {\ensuremath{{J\mskip -3mu/\mskip -2mu\psi\mskip 2mu}}\xspace}

\mathchardef\Upsilon="7107
\def\Y#1S{\ensuremath{\Upsilon{(#1S)}}\xspace}

\def\FourS {\Y4S}

\mathchardef\Deltares="7101
\mathchardef\Xi="7104
\mathchardef\Lambda="7103
\mathchardef\Sigma="7106
\mathchardef\Omega="710A

\def\Deltabar{\kern 0.25em\overline{\kern -0.25em \Deltares}{}\xspace}
\def\Lbar{\kern 0.2em\overline{\kern -0.2em\Lambda\kern 0.05em}\kern-0.05em{}\xspace}
\def\Sigbar{\kern 0.2em\overline{\kern -0.2em \Sigma}{}\xspace}
\def\Xibar{\kern 0.2em\overline{\kern -0.2em \Xi}{}\xspace}
\def\Obar{\kern 0.2em\overline{\kern -0.2em \Omega}{}\xspace}
\def\Nbar{\kern 0.2em\overline{\kern -0.2em N}{}\xspace}
\def\Xb{\kern 0.2em\overline{\kern -0.2em X}{}\xspace}

\def\X {\ensuremath{X}\xspace}

\def\BR         {{\ensuremath{\cal B}\xspace}}

\def\bpsikl     {\ensuremath{\Bz \to \jpsi \KL}\xspace}

\def\Bzbtox     {\ensuremath{\Bzb \to \X}\xspace}

\def\invfb   {\ensuremath{\mbox{\,fb}^{-1}}\xspace}

\def\mus  {\ensuremath{\rm \,\mus}\xspace}

\def\mus        {\ensuremath{\,\mu{\rm s}}\xspace}    

\def\degrees{\ensuremath{^{\circ}}\xspace}

\def\to                 {\ensuremath{\rightarrow}\xspace}

\def\pep2{PEP-II}
\def\BF{$B$ Factory}

\def\gsim{{~\raise.15em\hbox{$>$}\kern-.85em
          \lower.35em\hbox{$\sim$}~}\xspace}
\def\lsim{{~\raise.15em\hbox{$<$}\kern-.85em
          \lower.35em\hbox{$\sim$}~}\xspace}

\def\epstag  {\ensuremath{\varepsilon_{\rm tag}}\xspace}

\def\jetset74   {\mbox{\tt Jetset \hspace{-0.5em}7.\hspace{-0.2em}4}\xspace}

\usepackage{relsize}

\def\Mnu{\ensuremath{{\cal M}^2}}
\def\Mi{\ensuremath{{\cal M}_i^2}}
\def\Ms{\ensuremath{{\cal M}_s^2}}
\def\M{\ensuremath{{\cal M}_1^2}}
\def\MM{\ensuremath{{\cal M}_2^2}}

\def\fzz{\ensuremath{f_{00}}}
\def\Ns{\ensuremath{N_s}}
\def\Nd{\ensuremath{N_d}}

\begin{document}
{\pagestyle{empty}

\begin{flushright}
\babar-CONF-\BABARPubYear/\BABARConfNumber \\
SLAC-PUB-\SLACPubNumber \\
August 2004 \\
\end{flushright}

\par\vskip 5cm

\begin{center}
\Large \bf \boldmath Measurement of the Branching Fraction of
{\boldmath{$e^+e^- \rightarrow \BzBzb$ \\ at the $\Y4S$ Resonance}}
\end{center}
\bigskip

\begin{center}
\large The \babar\ Collaboration\\
\mbox{ }\\
\today
\end{center}
\bigskip \bigskip

\begin{center}
\large
\end{center}
 
We report a measurement of the branching fraction $e^+e^- \rightarrow \BzBzb$ 
with a data sample of 81.7~\invfb collected at the $\Y4S$ resonance with the \babar\
detector at the \pep2\ asymmetric-energy $e^+e^-$ storage ring.  Using
partial reconstruction of the decay $\Bzb \rightarrow D^{*+}
\ell^{-} \bar{\nu}_{\ell}$ we obtain a preliminary result of 
$\fzz = 0.486 \pm 0.010(stat.) \pm 0.009(sys.)$.
Our result does not depend on branching fractions of the $\Bzb$
and the $D^{*+}$ decay chains, on the simulated reconstruction efficiency, 
on the ratio of the charged and neutral $B$ meson lifetimes, nor 
on assumption of isospin symmetry.
This measurement is important for normalizing many $B$ decay branching fractions, 
and contributes to our understanding of isospin violation in the $\Upsilon(4S)$ system.


\vfill

\begin{center}

Submitted to the 32$^{\rm nd}$ International Conference on High-Energy Physics, ICHEP 04,\\
16 August-22 August 2004, Beijing, China

\end{center}

\vspace{1.0cm}
\begin{center}
{\em Stanford Linear Accelerator Center, Stanford University, 
Stanford, CA 94309} \\ \vspace{0.1cm}\hrule\vspace{0.1cm}
Work supported in part by Department of Energy contract DE-AC03-76SF00515.
\end{center}

\newpage
} 


\begin{center}
\small

The \babar\ Collaboration,
\bigskip

%
B.~Aubert,
R.~Barate,
D.~Boutigny,
F.~Couderc,
J.-M.~Gaillard,
A.~Hicheur,
Y.~Karyotakis,
J.~P.~Lees,
V.~Tisserand,
A.~Zghiche
\inst{Laboratoire de Physique des Particules, F-74941 Annecy-le-Vieux, France }
% ...

\end{center}\newpage

\section{INTRODUCTION}
\label{sec:Introduction}
Isospin violation in decays of $e^+e^- \rightarrow \BzBzb$ at 
the $\Y4S$ resonance results in a difference between the branching fractions
$\fzz \equiv {\cal B}(e^+e^- \rightarrow \BzBzb)$  
and
$f_{+-} \equiv {\cal B}(e^+e^- \rightarrow B^+ {B}^{-})$.
Measurements of the ratio $R^{+/0} \equiv f_{+-} / \fzz$,
summarized in Table~\ref{tab:R_measurements}, 
are consistent with unity within the errors~\cite{pdg2004}. 
Theoretical predictions for $R^{+/0}$ range from 1.03 to 1.25~\cite{eichten}.
Currently, almost all published measurements of $B$ meson branching
fractions make the assumption that $R^{+/0} = 1$.  
Precision measurements of $\fzz$, $f_{+-}$, and $R^{+/0}$ can be used
to eliminate this assumption and re-normalize all $B$ meson
branching fractions. 
\begin{table}[!htb]
\centering\caption{Summary of previous measurements of $R^{+/0}$}.
\begin{tabular}{lccc}\hline \hline
Decay $B\rightarrow$     & $\int{\cal L}dt$ & $R^{+/0}$ & Source \\ \hline
$J/\psi (K^{+}/K_{s}^{0})$
      & 81.9 fb$^{-1}$~~ & $1.006 \pm 0.036 \pm 0.031$  & \babar~\cite{haleh}  \\
$D^{*(+/0)}\ell\bar\nu$ 
      & 2.73 fb$^{-1}$  & $1.058 \pm 0.084 \pm 0.136$   & CLEO~\cite{godang02} \\
$J/\psi h^{(+/0)}$
      & 20.7 fb$^{-1}$   & $1.10 \pm 0.06 \pm 0.05$     & \babar~\cite{babar}\\
$J/\psi K^{*(+/0)}$  
      & 9.2 fb$^{-1}$   & $1.04 \pm 0.07 \pm 0.04$      & CLEO~\cite{silvia} \\ \hline \hline
\end{tabular}
\label{tab:R_measurements}
\end{table}


In this paper we report the first direct measurement of $\fzz$.  
The measurement is based on partial reconstruction of the decay 
$\Bzb \rightarrow D^{*+} \ell^{-}\bar{\nu}_{\ell}$ (the inclusion of charge-conjugate 
states is implied throughout this paper).
This allows a sizeable sample of double tagged events to be identified.
Comparison of the double-tag and the single-tag yields allows a determination of
$\fzz$ with minimal input from simulation.

The technique used to measure $\fzz$ is as follows:
in every event we reconstruct the decay $\Bzb \rightarrow D^{*+} \ell^{-}
\bar{\nu}_{\ell}$, as described further below.
The sample of events in which at least one $\Bzb \rightarrow D^{*+} \ell^{-}
\bar{\nu}_{\ell}$ candidate decay is found is labeled as ``single-tag sample''. 
The number of signal decays found in this sample is
\begin{equation}
\Ns =
      2 \NBB \fzz\, \epsilon_{s} \, 
      {\cal B}(\Bzb \rightarrow D^{*+} \ell^- \bar{\nu}_{\ell}),
\label{eq:ns}
\end{equation}
where $\NBB = (88726 \pm 23)\times 10^3$ is the total number of $\BB$
events in the data sample and $\epsilon_{s}$ is the reconstruction
efficiency of the decay $\Bzb \rightarrow D^{*+} \ell^{-}
\bar{\nu}_{\ell}$.  The technique for measuring \NBB\ is 
described in~\cite{Aubert:2002hc}.
The data sample has a mean energy of 10.580~\gev~\cite{beam_spread} and
an energy spread of only $4.6~MeV$. Such a small spread means that any energy
dependence of \fzz\ has a negligible effect on the central value.
The subset of single-tag events in which two $\Bzb \rightarrow D^{*+} 
\ell^{-}\bar{\nu}_{\ell}$ candidates are found is labeled as ``double-tag sample''. 
The number of such events is
\begin{equation}
\Nd =  
      \NBB \, \fzz\, \epsilon_{d} \,
      [{\cal B}(\Bzb \rightarrow D^{*+} \ell^- \bar{\nu}_{\ell})]^2 ,  
\label{eq:nd}
\end{equation} 
where $\epsilon_{d}$ is the efficiency to reconstruct two $\Bzb \rightarrow D^{*+}
\ell^{-}\bar{\nu}_{\ell}$ decays in the same event.
Note that every double-tag event contributes two entries to the
single-tag sample.
Using Eq.~(\ref{eq:ns}) and Eq.~(\ref{eq:nd}), the ratio \fzz\ is given by
\begin{equation}
\fzz =  
       {C N_{s}^{2} \over {4 N_{d} \NBB} },
\label{eq:f00}
\end{equation}
where we have defined the coefficient $C \equiv \epsilon_{d} / \epsilon_{s}^{2}$. 
$C=1$ if the efficiencies for detecting each $B$ meson are uncorrelated in 
double-tag events.

\newpage     

\section{THE \babar\ DETECTOR AND DATASET}
\label{sec:babar}

The \babar\ data sample used in this paper consists of 81.7~\invfb collected 
at the $\Y4S$ resonance (the on-resonance sample) and 9.6~\invfb collected 40~MeV 
below the resonance (the off-resonance sample).  
Simulated $\BB$ events were analyzed through the same
analysis chain as the data. The equivalent luminosity of the
simulated sample is approximately three times that of the on-resonance data. 

A detailed description of the \babar\ detector and the algorithms used
for track reconstruction and particle identification is provided
elsewhere~\cite{babar_nim}. A brief summary is given here.
High-momentum particles are reconstructed by
matching hits in the silicon vertex tracker (SVT) with track elements
in the drift chamber (DCH). Lower momentum tracks, which do not leave
signals on many wires in the DCH due to the bending induced by a
magnetic field, are reconstructed by the SVT alone.
Electrons are identified with the ratio of the track momentum to the
associated energy deposited in the calorimeter (EMC), the transverse
profile of the shower, the energy loss in the drift chamber, and the
information from a Cherenkov detector (DIRC).  
Muons are identified in the instrumented flux return (IFR), composed
of resistive plate chambers and layers of iron.
Muon candidates are required to have a path length and hit
distribution in the instrumented flux return and energy deposition in
the EMC consistent with that expected for a minimum-ionizing particle.
The Cherenkov light emission in the DIRC is then employed to further
reject kaons misidentified as muons by requiring muon candidates to
have a kaon hypothesis probability less than 5\%.  

Hadronic events are selected by requiring at
least four charged particle tracks reconstructed by the silicon
vertex detector and the drift chamber. 
To reduce background from continuum $\epem\to\qqbar$, where $q$ stands
for a $u$, $d$, $s$, or $c$ quark, the ratio $R_2=H_2/H_0$ of the
second to the zeroth Fox-Wolfram moments is used~\cite{wolfram}.

\section{ANALYSIS METHOD}
\label{sec:Analysis}

We reconstruct the decays $\Bzb \rightarrow D^{*+} 
\ell^{-} \bar{\nu}_{\ell}$ with a partial reconstruction technique.  
The application of the technique to this mode was first proposed by 
the ARGUS Collaboration~\cite{argus} and has been used by CLEO~\cite{godang02}, 
DELPHI~\cite{delphi}, OPAL~\cite{opal}, and \babar~\cite{franco}.  
In this technique, only the lepton from the decay $\Bzb \rightarrow D^{*+} 
\ell^{-} \bar{\nu}_{\ell}$ and the soft pion from the decay
$D^{*+}\to \Dz \pi^+$ are used. No attempt is made to reconstruct
the $\Dz$, resulting in high reconstruction efficiency. 

To suppress leptons from charm decays, all lepton candidates
(electrons and muons) are required to have 
momentum between 1.5~\gevc and 2.5~\gevc 
in the $\epem$ center-of-mass (CM) frame.
Soft pion candidates are required to have CM momentum
between 60~\mevc and 200\mevc.
As a consequence of the limited phase space available in the $D^{*+}$
decay, the soft pion is emitted within a one radian-wide cone centered
about the $D^{*+}$ direction in the CM frame.  The $D^{*+}$
four-momentum can therefore be computed by approximating its direction
as that of the soft pion, and parameterizing its momentum as a linear
function of the soft pion momentum, with parameters obtained from the
simulation. 
The presence of an undetected neutrino is inferred from conservation
of momentum and energy. The neutrino invariant mass squared is calculated:
\begin{eqnarray}
\Mnu \equiv (E_{\mbox{\rm beam}}-E_{{D^*}} - 
E_{\ell})^2-({\bf{p}}_{{D^*}} + {\bf{p}}_{\ell})^2\ ,
\label{eqn:mms}
\end{eqnarray}
where $E_{\mbox{\rm beam}}$ is the beam energy and $E_{\ell}~(E_{{D^*}})$ 
and ${\bf{p}}_{\ell}~({\bf{p}}_{{D^*}})$ are the CM energy and momentum 
of the lepton (the $D^*$ meson). 
If the decay is properly reconstructed and the neutrino is the only
missing particle, the $\Mnu$ distribution will peak
near zero for signal events. Background events, however, are
spread over a wide range of $\Mnu$ values.


In what follows, we use the symbol $\Ms$ to denote $\Mnu$ for any
candidate in the single-tag sample.
In the double-tag sample, we randomly choose one of the two
reconstructed $\Bzb \rightarrow D^{*+} \ell^- \nu_l$ candidates 
as ``first'' and the other as ``second''. Their $\Mnu$ values are 
labeled $\M$ and $\MM$, respectively.
For each of the variables $\Mi$ ($i=s,1,2$), we define a signal region 
$\Mi > -2~$GeV$^{2}/c^4$ and the sideband $-8 < \Mi < -4~$GeV$^{2}/c^4$.


In addition to signal $\Bzb \rightarrow D^{*+} \ell^- \nu_l$ decays, 
the single-tag and double-tag samples contain several types of events: 
\begin{itemize}
\item Continuum $\epem\to\qqbar$ background.
\item Combinatorial $\BB$ background, formed from random combinations of
reconstructed leptons and soft pions.  This background can also be due to the 
low momentum soft pions not coming from a $D^{*}$, produced by either the same $B$ 
or other $B$~\cite{e791}.

\item Peaking $\BB$ background, composed of $\Bb \to D^{*} (n\pi)
\ell \bar{\nu}_{\ell}$ decays with or without an excited charmed
resonance ($D^{**}$)~\cite{e961}, where the reconstructed soft
pion comes from the decay $D^{*+}\to \Dz \pi^+$, leading to an
accumulation of these events at high values of $\Mi$.
These events are peaking background and are produced both
by $\Bzb$ and $B^-$ decays.  Their $\Mi$ distribution differs from
the signal, which allows us to extract their contribution in a fit. 
Such events are suppressed by the requirement $p_\ell > 1.5
\gevc$ on the lepton CM momentum.
\item The decays $\Bzb \to D^{*+} \tau^- \bar{\nu}_\tau$ 
and $\Bzb \to D^{*+} \ell^{-}\bar{\nu}_{\ell} (n\gamma)$ 
may be used for the measurement of \fzz\ and 
are therefore considered as signal.  These events peak in $\Mi$ 
and come from $\Bzb$ decays.

\end{itemize}
The double-tag sample contains two additional types of background:
\M-combinatorial and \M-peaking.
In \M-combinatorial (\M-peaking) background events the first candidate
is combinatorial (peaking) background.


To determine \Ns\ and \Nd, we perform binned $\chi^2$ fits to 
one-dimensional histograms of the \Ms\ and \MM\ distributions of
on-resonance data events, ranging from $-8$ to $2~$GeV$^{2}/c^4$.
Before fitting, we subtract the continuum background contribution from
the histograms. This is done using the \Ms\ and \MM\ distributions of
off-resonance data, scaled to account for the ratio of on-resonance to
off-resonance luminosities and the CM energy dependence of the
continuum production cross-section.
In addition, the contributions of the \M-combinatorial and \M-peaking
backgrounds are subtracted from the \MM\ histogram before the fit.
The \M-combinatorial background is determined from the \M\ sideband,
which contain only continuum and combinatorial background events.  
This histogram is scaled by the ratio of the number of combinatorial 
events in the signal region and the sideband, determined from the simulation.
The \M-peaking background subtraction is based on the simulated
\M-peaking events.

After the subtraction, the \Ms\ and \MM\ histograms are fit
separately, using a function whose value for bin $j$ of the histogram
is
\begin{equation}
f_j = \sum_t N^t P^t_j,
\end{equation}
where $N^t$ is the number of events of type $t$ ($t=$ signal,
combinatorial, peaking) populating the histogram, and $P^t_j$ is the
bin $j$ value of a discrete probability density function (PDF) obtained from
simulated events of type $t$, normalized such that 
$\sum_j P^t_j = 1$. 
The fit determines the parameters $N^t$ by minimizing:
\begin{equation}
\chi^2 = \sum_j{{(H_j - f_j)^2 \over \sigma_{H_j}^2 + \sigma_{f_j}^2}},
\label{eq:fitchi2}
\end{equation}
where $H_j$ is the number of entries in bin $j$ of the histogram being
fit; $\sigma_{H_j}$ is the statistical error on $H_j$, including
uncertainties due to the background subtractions described above; and
$\sigma_{f_j}$ is the error on $f_j$, determined from the errors on
$P_j^t$, which are due to the finite size of the simulated sample.

\begin{figure}[!htb]
\begin{center}
\vspace*{-2.7cm}
\hspace{-0.5cm}
\includegraphics[width=11.2cm]{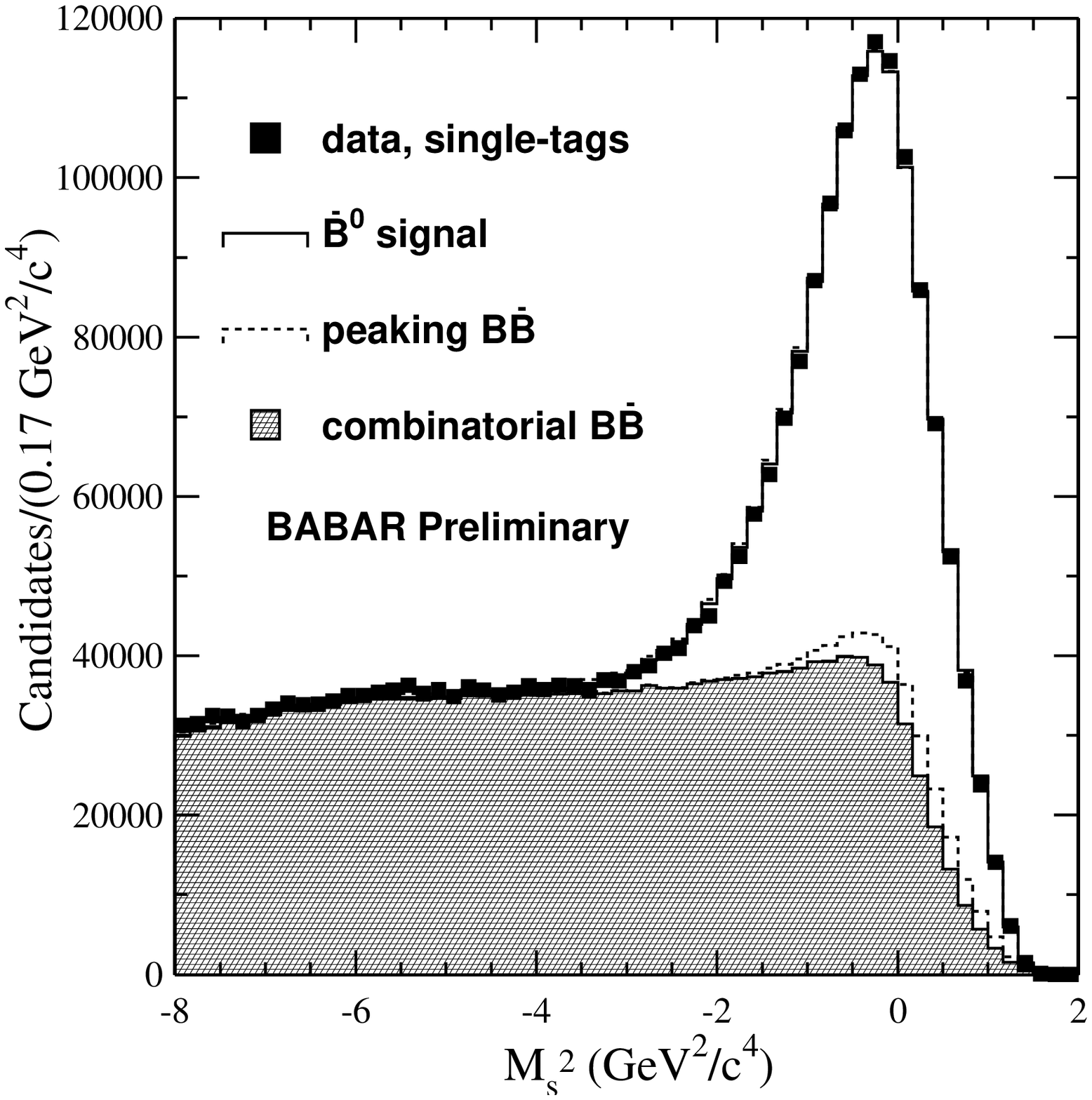}  
\vspace*{-5.1cm}
\hspace{-0.5cm}
\includegraphics[width=11.2cm]{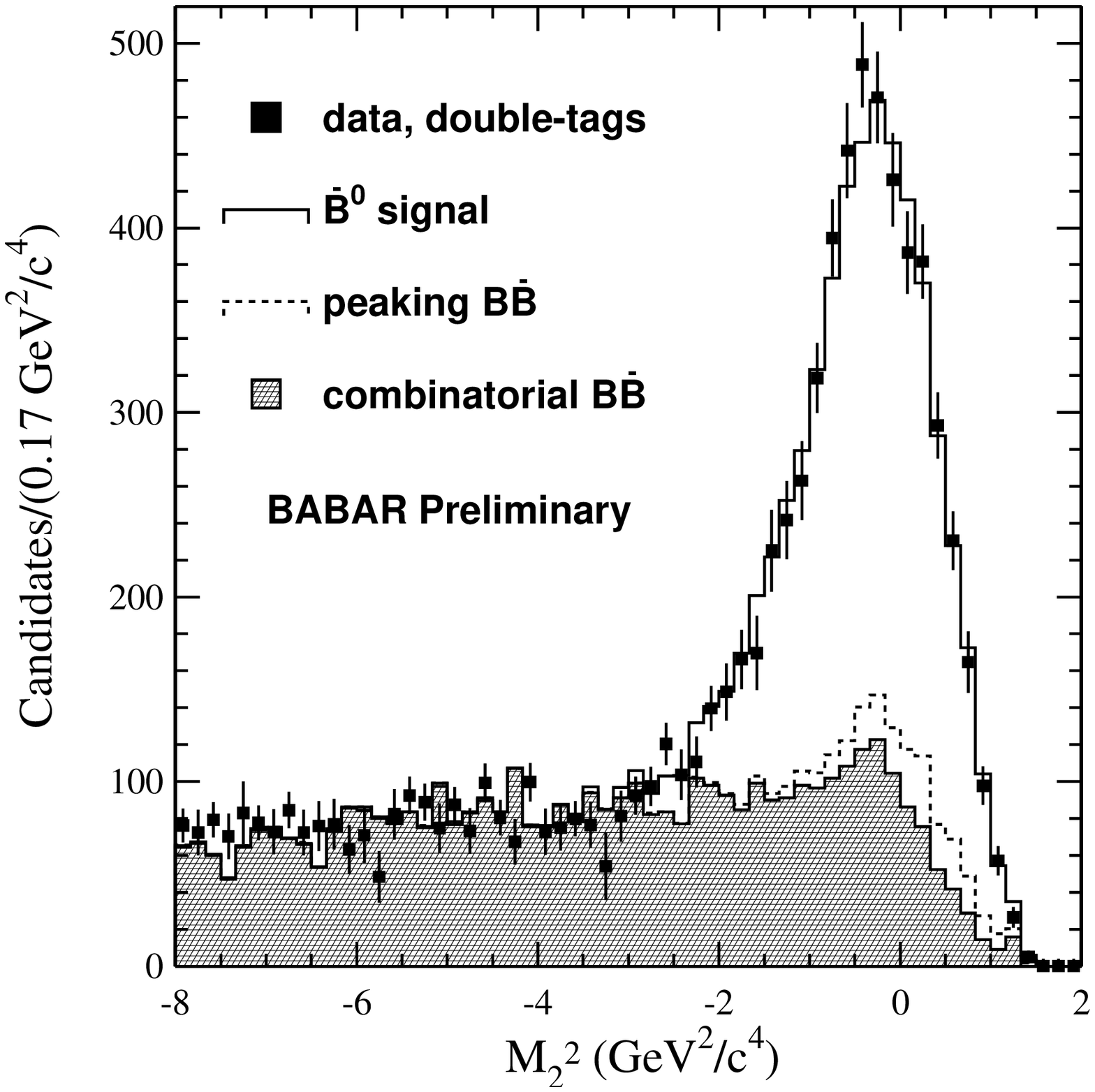}  
\vspace*{-1.5cm}
\caption{The \Ms\ (top) and \MM\ (bottom) distributions of the
on-resonance samples.  The continuum background has been subtracted from
the \Ms\ distribution. For the \MM\ distribution, the \M-combinatorial, and 
the \M-peaking have been subtracted.
The levels of the simulated signal, peaking \BB\ and
combinatorial \BB\ background contributions are obtained from the
fit.}
\label{fig:rightsign}
\end{center}
\end{figure}

The results of the fits are presented in Table~\ref{tab:numbers}.
The \Ms\ and \MM\ distributions are shown in
Fig.~\ref{fig:rightsign}, with the
contributions of the different event types indicated.
The fits yield the values $\Ns = 786300 \pm 1950$ and
$\Nd = 3560 \pm 80$. 
Using the simulation we determine $C = 0.9946 \pm 0.0078$, 
where the error is due to the finite size of the simulated sample.
Eq.~(\ref{eq:f00}) then gives
$\fzz = 0.486 \pm 0.010$, where the error is due to data statistics only.
\begin{table}[!htb]
\caption{Numbers of entries of different types 
found by the fits to the \Ms\ and \MM\ histograms in the signal region.
Also shown are the numbers of entries of subtracted backgrounds
and the confidence levels of the fits.}
\begin{center}
\begin{tabular}{lcc} 
\hline \hline
Source                      & \Ms\               & \MM\           \\
\hline 
{\raisebox{-0.4ex}{Combinatorial $\BB$}}         & $558090 \pm 760$   & $1520 \pm 40$  \\
Peaking $\BB$               & $68170  \pm 260$   & $300  \pm 20$  \\
Signal                      & $786300 \pm 2000$  & $3560 \pm 80$  \\ 
\hline
Continuum                   & $238500 \pm 1300$  & $160 \pm 40$   \\
$\M$-combinatorial          & ---                & $180 \pm 20$   \\
$\M$-peaking                & ---                & $60  \pm 10$   \\
\hline 
$\chi^2/$d.o.f.              & $41/56$            & $48/56$        \\
Confidence level            & 93\%               & 77\%            \\
\hline \hline 
\end{tabular}
\end{center}
\label{tab:numbers}
\end{table}

\begin{figure}[!htb]
\begin{center}
\vspace*{-2.3cm}
\hspace*{-0.6cm}
\includegraphics[width=18cm]{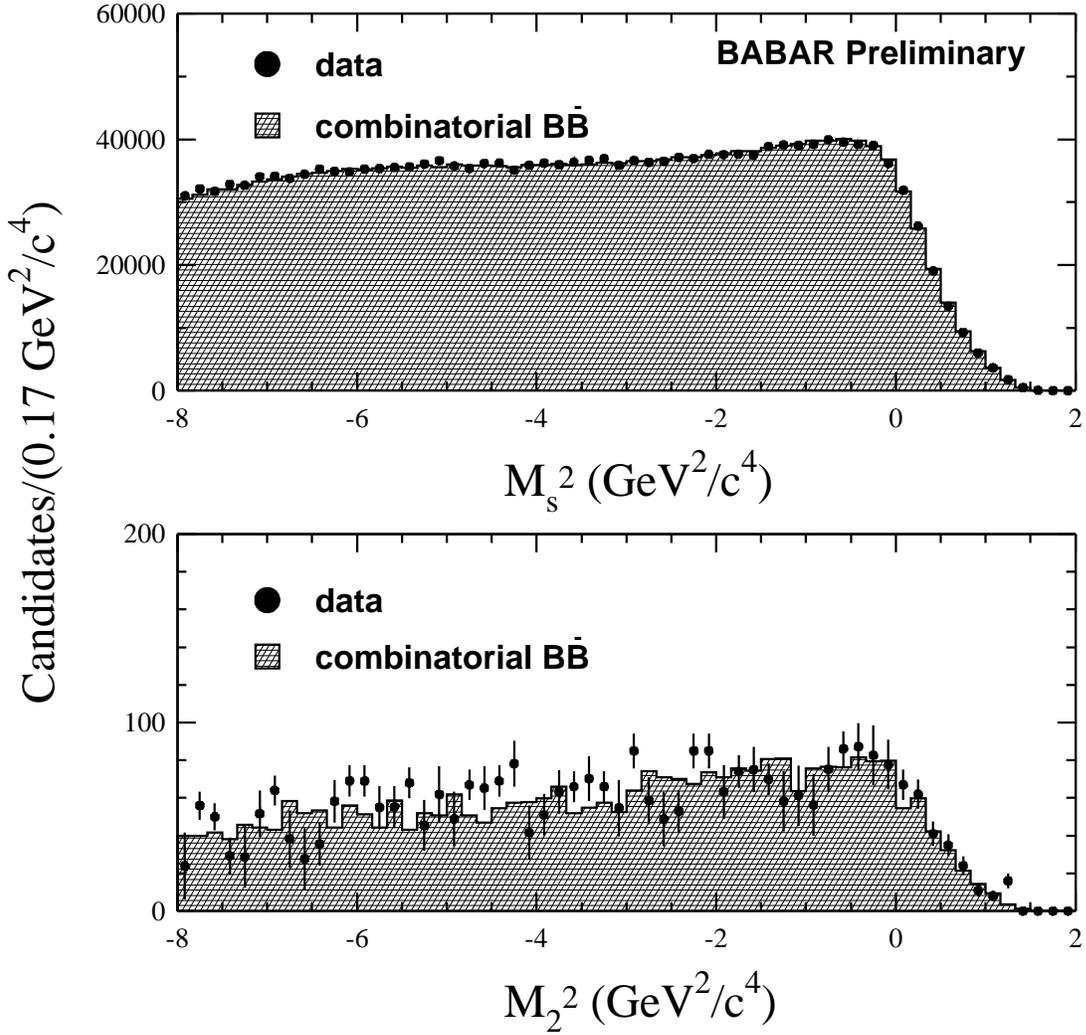}
\vspace*{-2.5cm}
\caption{The \Ms\ (top) and \MM\ (bottom) 
distributions of the same-charge on-resonance samples.
The continuum background has been subtracted from the \Ms\ distribution.
For the \MM\ distribution, the continuum background, 
the \M-combinatorial and the \M-peaking backgrounds have been subtracted.
The level of the simulated combinatorial \BB\ background is 
obtained from the fit.}
\label{fig:wrongsign}
\end{center}
\end{figure}
\begin{figure}[!htb]
\begin{center}
\vspace*{-1.0cm}
\begin{tabular}{lrlr} \hspace{-1.0cm}
\includegraphics[height=11.5cm,width=9.5cm]{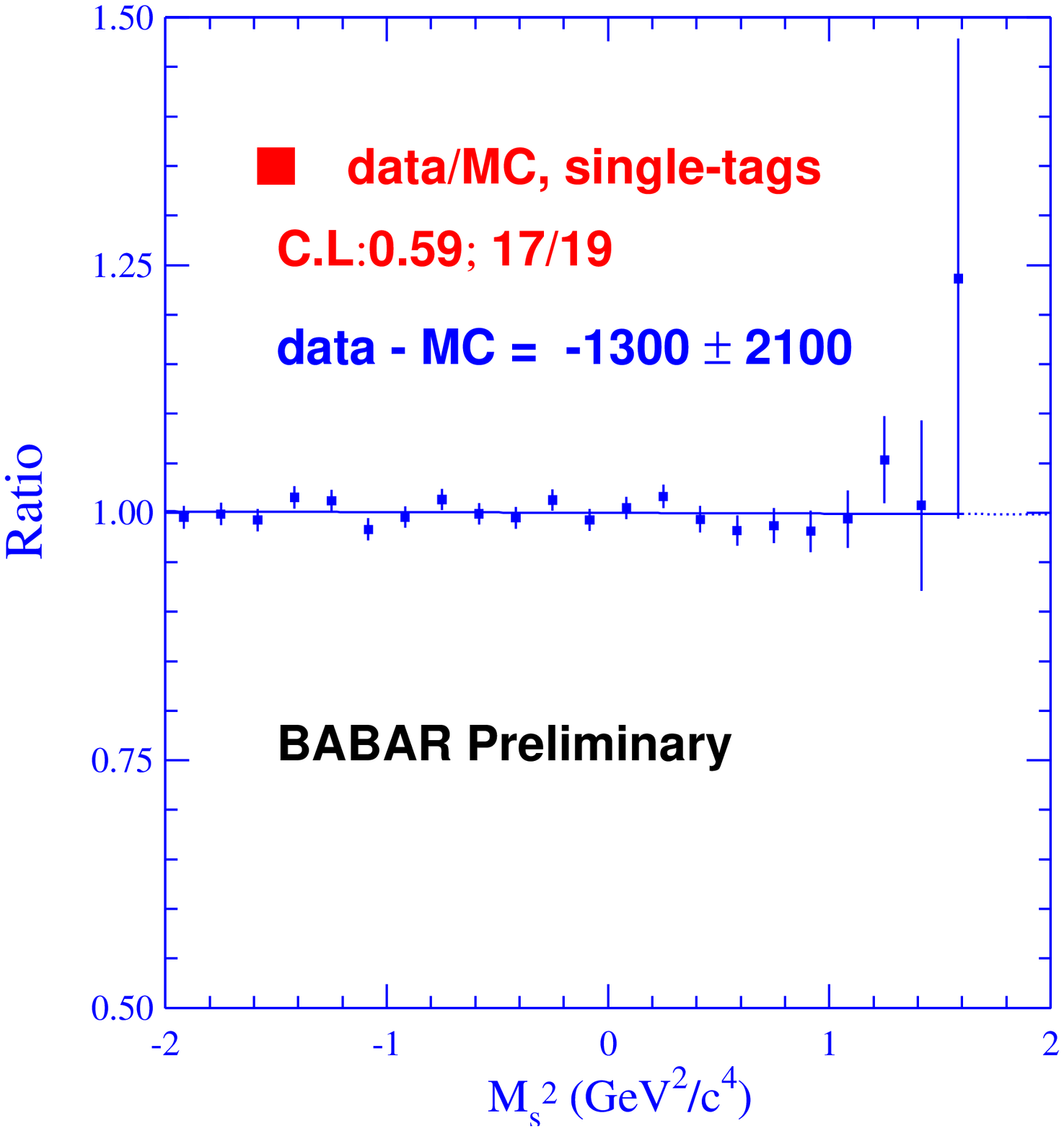}& \hspace{-1.5cm}  
\includegraphics[height=11.5cm,width=9.5cm]{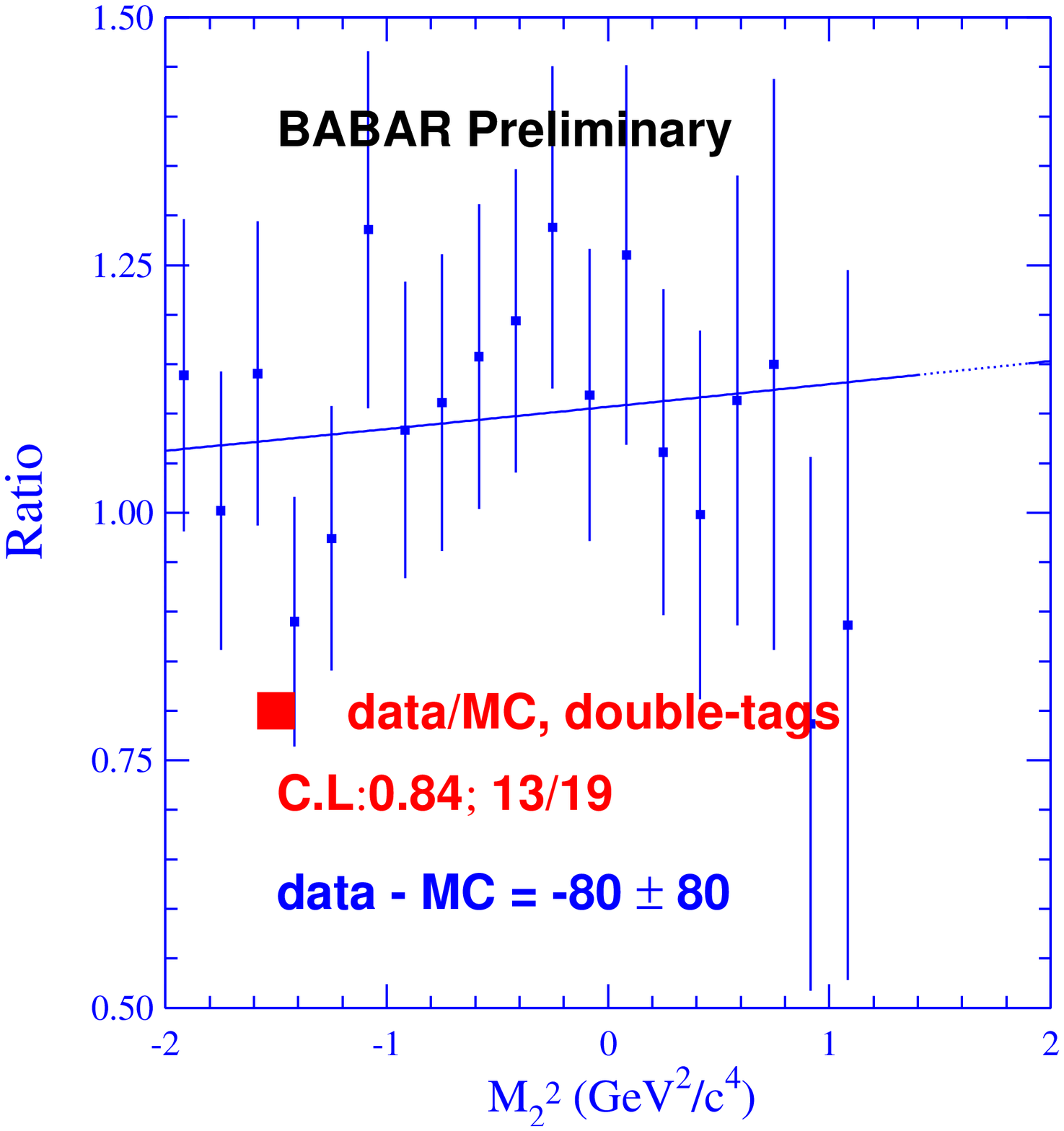}& \vspace{-2.5cm} \\ 
\hspace{-1.0cm}
\includegraphics[height=11.5cm,width=9.5cm]{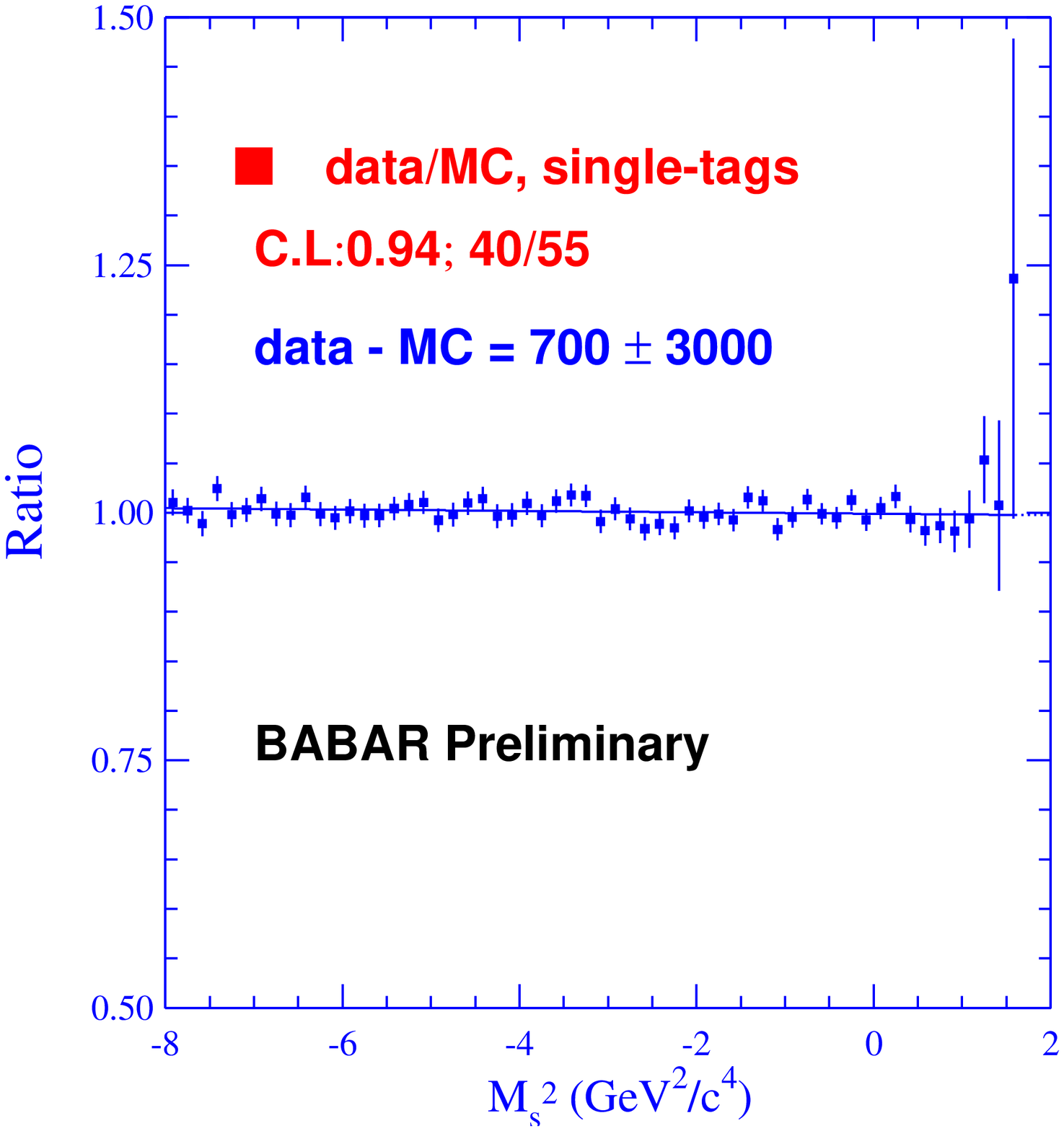}& \hspace{-1.5cm}
\includegraphics[height=11.5cm,width=9.5cm]{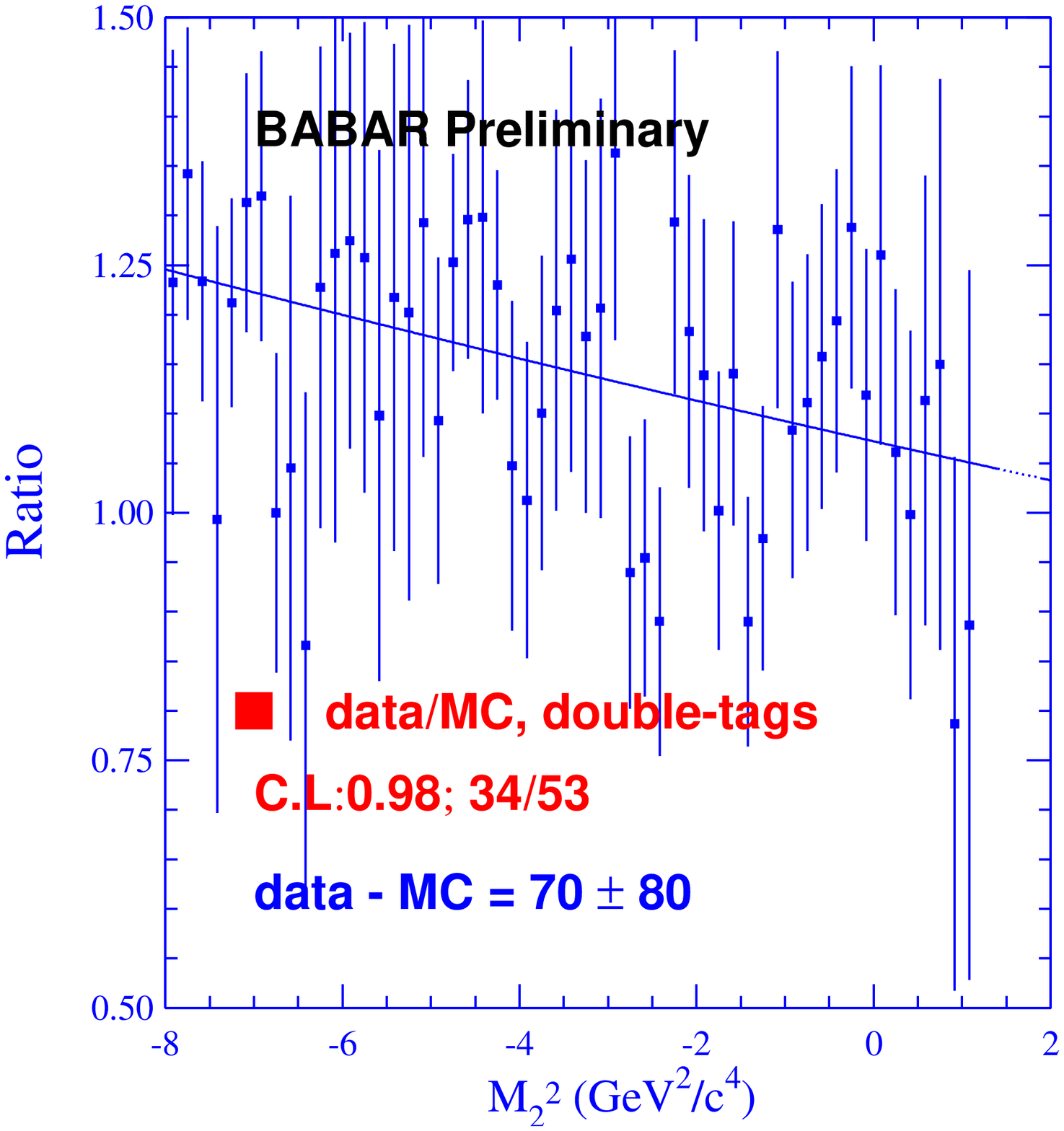}
\end{tabular}
\vspace*{-0.5cm}
\caption {The ratio between data and the combinatorial \BB\ background of the
same-charge sample both for the \Ms\ and \MM\ summed over the signal region 
and over all bins. The values are fit to a constant.
upper left: for the \Ms\ summed over the signal region;
upper right: for the \MM\ summed over the signal region;
lower left: for the \Ms\ summed over all bins;
lower right: for the \MM\ summed over all bins.}
\label{fig:chi_ratio}
\end{center}
\end{figure}

To determine how well the simulation reproduces the \Ms\ and \MM\
distributions of the combinatorial background in the data,
we study the distributions of a sample of same-charge
candidates, in which the lepton and soft pion have the same
electric charge. This sample contains only continuum and 
combinatorial \BB\ background.
We fit the continuum-subtracted \Ms\ and \MM\ histograms of the
same-charge sample using the function $f'_j = N P'_j$, where $P'_j$ is
the bin $j$ value of the PDF of
same-charge simulated \BB\ events, normalized such that $\sum_j P'_j =
1$, and the parameter $N$ is determined by the fit.
The histograms, overlaid with the fit function, are shown in
Fig.~\ref{fig:wrongsign}.  The ratio between these two histograms is 
fitted to a constant both for the \Ms\ and \MM\ summed over the signal 
region and over all bins are shown in Fig.~\ref{fig:chi_ratio}.
The accumulated differences 
$D \equiv \sum_j (H'_j - f'_j)$
between the same-charge data histograms $H'_j$ and the fit
functions are summarized in Table~\ref{tab:ws_ratio}. 
Their consistency with zero indicates that the distributions of simulated
combinatorial \BB\ background events do not lead to significant fake
signal yields.
\begin{table}[!htb]
\centering\caption{The difference $D \equiv \sum_j (H'_j - f'_j)$ between the
same-charge data histogram and the fit function, summed over all bins
or over the signal region only. Also shown are the fit $\chi^2$/d.o.f. values
and confidence levels.}
\vspace{2mm}
\begin{tabular}{l|cc|cc}  
\hline \hline 
          &    \multicolumn{2}{c|}{Signal region} 
			& \multicolumn{2}{c}{All bins} \\
          &        \Ms        & \MM    &   \Ms        & \MM      \\ 
\hline
$D$             & $-1300 \pm 2100$     & $-80 \pm 80$  
	        & $-700 \pm 3000$      & $-70 \pm 80$   \\
$\chi^2$/d.o.f. & 17/19  & 13/19 & 40/55  & 34/53           \\ 
C.L.($\%$)      & 59     & 84    & 94     & 98 \\
\hline \hline
\end{tabular}
\label{tab:ws_ratio}
\end{table}

\section{SYSTEMATIC STUDIES}
\label{sec:Systematics}
We consider several sources of systematic uncertainties in $\fzz$. 
All estimated errors are an absolute systematic uncertainties in $\fzz$ 
and summarized in Table~\ref{tab:sys_error}.
\begin{table}[!htb]
\caption{Summary of the absolute systematic errors for $f_{00}$.}
\begin{center}
\begin{tabular}{lc} \hline \hline
Source                                & $\delta(f_{00})$   \\ \hline
\M-combinatorial                      & $0.0005$           \\
\M-peaking                            & $0.0005$           \\
Same charged events                   & $0.0025$           \\
Peaking background                    & $0.004$            \\      
$B$-meson counting                    & $0.0055$           \\
$\Upsilon(4S) \to$ non-$\BB$          & $0.0025$           \\
Efficiency correlation                & $0.004$            \\ 
Monte Carlo statistics                & $0.002$            \\  \hline
Total                                 & $0.009$            \\  \hline \hline
\end{tabular}
\end{center}
\label{tab:sys_error}
\end{table}   

\begin{enumerate}
\item The systematic uncertainty from the \M-combinatorial contribution subtraction 
in the \MM\ histogram is 0.0005.  The error is obtained by varying the total
\M-combinatorial background by its statistical error and repeating the analysis.
\item An error of 0.0005 is estimated due to the subtraction of the
\M-peaking contribution in the \MM\ histogram. The error is obtained 
by comparing the ratio between the numbers of subtracted \M-peaking 
and \M-combinatorial events with their ratio of peaking and
combinatorial events in Table~\ref{tab:numbers}.
\item Propagating the errors on the quantities $D$ (same charged events) 
of Table~\ref{tab:ws_ratio} leads to an error of 0.0025 on \fzz.  
To determine this error we vary the signal events both for the single-tag 
and the double-tag samples. 
The largest uncertainty then is taken for the uncertainty on $f_{00}$.
\item The PDFs $P^t$ ($t=$ peak) of the peaking background come from
simulated event samples containing different $D^{**}$ resonances or 
non-resonant events.  We vary the ratio of the branching fraction of
the resonant and the non-resonant production such that the variation 
of this ratio is wide enough to include poorly known decays.  
We repeat the analysis procedure to determine \Ns\ and \Nd.  
The resulting error on \fzz\ is 0.004. 
\item Uncertainties in the branching fractions of $\Bzb \to D^{*+} \tau^-
\bar{\nu}_\tau$ and $\Bzb \to D^{*+} \ell^{-}\bar{\nu}_{\ell} 
(n\gamma)$ relative to $\Bzb \to D^{*+} \ell^{-}\bar{\nu}_{\ell}$ 
lead to uncertainties in the PDFs $P^t$ ($t=$ signal) of the signal events. 
This uncertainty in \fzz\ is negligible.
\item The error due to the uncertainty in \NBB\ is 0.0055.
It includes the uncertainties for differences in the cross sections
and efficiencies for muon pairs and continuum events between
on-resonance and off-resonance samples, hadronic selection criteria
and the uncertainties in the tracking efficiency.
\item In this paper the impact of non-$\BB$ decays of the $\Upsilon(4S)$ 
on $B$-meson counting has been accounted for as a systematic error.
The upper limit for the branching fraction of $\Upsilon(4S)$ decays into 
non-$\BB$ is $4\%$ at $95\%$ confidence level~\cite{nonbbar}.
We conservatively estimate the systematic error by decreasing the $4\%$ 
upper limit on the branching fraction to $2\%$.
From this variation we estimate an error of
0.0025 due to the effect on \NBB\ of a possible decay. 
\item We note that the lepton momentum spectrum in the Monte Carlo simulation 
is different from the one we observe in the data.  We tune the simulation to the 
data by rejecting simulated events in such a way that the two lepton momentum 
spectra agree.  We repeat the analysis procedure without the rejected events.
The systematic error due to the uncertainty in the lepton momentum spectrum
is negligible.
\item There is a small efficiency correlation between the single-tag and 
the double-tag samples.
The systematic uncertainty due to this efficiency correlation is estimated 
by propagating the Monte Carlo simulation systematics error of $C$ 
into $f_{00}$.  The simulation statistical error in $C$ leads 
to a 0.004 error in \fzz.  
In addition to the Monte Carlo simulation systematics error of $C$,
we study the effect of track multiplicity on the efficiency correlation.
\item We perform a similar procedure as mentioned above for the pion momentum 
spectrum. The error due to the uncertainty in the pion momentum spectrum is 
negligible.
\item An error of 0.002 is due to the finite size of the simulated
sample, calculated using $\sigma_{f_j}$ in
Eq.~(\ref{eq:fitchi2}).
\item The $\chi^2$ estimator used in Eq.~(\ref{eq:fitchi2}) can be biased.  
We did an alternative binned likelihood fit and found that the result differed by
only $0.03\%$ for \fzz.
\end{enumerate}

We combine the uncertainties given above in quadrature to determine an
absolute systematic error of 0.009 in $\fzz$.


\section{SUMMARY}
\label{sec:Summary}
To summarize, using partial reconstruction of the decay $\Bzb \rightarrow 
D^{*+} \ell^- \nu_l$ we have obtained a preliminary result for the branching fraction
\begin{equation}
\fzz = 0.486 \pm 0.010 \pm 0.009, 
\end{equation}
where the first error is statistical and the second is systematic.
Since this measurement is done by comparing the numbers of events with
one and two reconstructed $\Bzb \rightarrow D^{*+} \ell^- \nu_l$ decays, 
it does not depend on branching fractions of the $\Bzb$
and the $D^{*+}$ decay chains, on the simulated reconstruction efficiency, 
on the ratio of the charged and neutral $B$ meson lifetimes, nor on assumptions 
of isospin symmetry.

\section{ACKNOWLEDGMENTS}
\label{sec:Acknowledgments}

We are grateful for the 
extraordinary contributions of our \pep2\ colleagues in
achieving the excellent luminosity and machine conditions
that have made this work possible.
The success of this project also relies critically on the 
expertise and dedication of the computing organizations that 
support \babar.
The collaborating institutions wish to thank 
SLAC for its support and the kind hospitality extended to them. 
This work is supported by the
US Department of Energy
and National Science Foundation, the
Natural Sciences and Engineering Research Council (Canada),
Institute of High Energy Physics (China), the
Commissariat \`a l'Energie Atomique and
Institut National de Physique Nucl\'eaire et de Physique des Particules
(France), the
Bundesministerium f\"ur Bildung und Forschung and
Deutsche Forschungsgemeinschaft
(Germany), the
Istituto Nazionale di Fisica Nucleare (Italy),
the Foundation for Fundamental Research on Matter (The Netherlands),
the Research Council of Norway, the
Ministry of Science and Technology of the Russian Federation, and the
Particle Physics and Astronomy Research Council (United Kingdom). 
Individuals have received support from 
CONACyT (Mexico),
the A. P. Sloan Foundation, 
the Research Corporation,
and the Alexander von Humboldt Foundation.

\end{document}